\documentclass[aps,prl,reprint]{revtex4-2}
\usepackage[utf8]{inputenc}
\usepackage[tbtags]{amsmath}
\usepackage{amssymb}
\usepackage{graphicx}
\usepackage{mathtools}
\usepackage{dsfont}
\usepackage{enumitem}
\usepackage{bbm}
\usepackage{hyperref}
\usepackage{cases}
\usepackage[usenames,dvipsnames]{color}
\usepackage[sort&compress]{natbib}

\begin{document}
\title{An explicit example for the high temperature convolution: crossover between the binomial law $B(2,1/2)$ and the arcsine law}
\author{Pierre Mergny}
\email{mergny.pierre@gmail.com}
\affiliation{LPTMS,  CNRS,  Univ.   Paris-Sud,  Université  Paris-Saclay,  91405  Orsay,  France}
\affiliation{Chair of Econophysics $\&$ Complex Systems, Ecole Polytechnique, 91128 Palaiseau Cedex, France}

\begin{abstract}
 In this note, we  study  the high-temperature convolution introduced in Ref.\  \cite{mergny_cconv},  between two symmetric Bernoulli distributions. We give an analytical expression for both the Stieltjes transform and the density. This result provides the first non-trivial expression for the high-temperature convolution of two distributions and  gives a new family of densities, interpolating between the centered binomial distribution with number of trials $n{=}2$ and probability of success $p{=}1/2$, and the centered and re-scaled arcsine law.  
\end{abstract}

\maketitle

\section{Introduction}
Understanding the spectrum of a large random matrix  is of special interest in  fields as diverse as high energy physics  \cite{wigner1951statistical,szabo2010chern}, statistics \cite{Wishart},  disordered systems \cite{Edwards_1975,sherrington1975solvable}, finance \cite{bouchaud2000theory}, economy \cite{moran2019may}, ecology \cite{may72} and genetics \cite{amir,guo2020stability} to cite a few. A particular - yet fundamental - case  of interest concerns the situation where one is dealing with the \emph{sum} of random matrices. 

{\bf Classical and free convolution -}
Important progress under this setting  has been made in the 90s with the introduction of \emph{free probability} \cite{Voiculescu1991,Voiculescu1997free,Mingo2017free}.  To fix things, we consider $\mathbf{a}, \mathbf{b} \in \mathbb{R}^N$ such that as $N$ goes  to infinity,  $\mu_{\mathbf{a}} := (1/N) \sum_{i=1}^N \delta_{a_i} \to \mu_{A}$ and similarly $\mu_{\mathbf{b}} \to \mu_B$, and  a random rotation matrix $\mathbf{O} \sim \mathrm{Haar}(\mathsf{O}(N))$, where $\mathsf{O}(N)$ is the group of $(N \times N)$ orthogonal matrices. As $N$ goes to infinity, the spectrum of $\mathrm{Diag}(\mathbf{a}) + \mathbf{O} \mathrm{Diag}(\mathbf{b})  \mathbf{O}^{\mathsf{T}}$  only depends on the limiting distributions  $\mu_A$ and $\mu_B$, and is known as the \emph{free convolution} of the former distributions, denoted by  $\mu_A \boxplus \mu_B$. In practice, one can compute the distribution $\mu_A \boxplus \mu_B$ thanks to the so-called \emph{R-transform}, and we refer to the textbook \cite{potters-bouchaud} for further details. The two symmetric matrices $ \mathbf
{A} := \mathrm{Diag}(\mathbf{a})$ and $\mathbf{B} := \mathbf{O} \mathrm{Diag}(\mathbf{b})  \mathbf{O}^{\mathsf{T}}$ are said to be (asymptotically) \emph{free} or equivalently their eigenbasis are in “generic positions”. Another way to view free probability is to notice that $ \mathbf
{A}$ and $\mathbf{B}$ are  “maximally non-commutative”  self-adjoint objects, in the sense that if both $\mu_A$ and $\mu_B$ have zero mean and if we denote by $\tau(.) := \mathrm{Tr}(.)/N$, we have at large $N$, $\tau( \mathbf{A}^2 \mathbf{B}^2) = \tau(\mathbf{A}^2) \tau(\mathbf{B}^2) =m_2[\mu_A] m_2 [\mu_B]$ with $m_2[\mu] := \int x^2 \mathrm{d}\mu $ but  $\tau(\mathbf{A} \mathbf{B} \mathbf{A} \mathbf{B}) = 0 \neq \tau(\mathbf{A}^2) \tau(\mathbf{B}^2)$. From a purely combinatorial point of view \cite{Nica2006}, this has a simple interpretation: the former corresponds to a non-crossing partition, while the latter corresponds to a crossing one. As $N \to \infty$, only non-crossing terms contribute to the computation of the moments of $\mathbf{A}+\mathbf{B}$,  in the free probability setting.

On the contrary, if now one is looking at the spectrum of $\mathrm{Diag}(\mathbf{a}) + \mathbf{P} \mathrm{Diag}(\mathbf{b})  \mathbf{P}^{\mathsf{T}}$ where $\mathbf{P} \sim \mathrm{Unif}( \mathsf{S}(N))$ is a (uniform) random permutation matrix, one is summing diagonal matrices and the limiting spectral distribution for the sum is now simply given by the classical convolution $\mu_A \ast \mu_B$. Thus, classical convolution naturally appears in Random Matrix Theory (RMT)  when the eigenbasis of both symmetric matrices are perfectly aligned, or said differently when one is looking at the spectrum of large commutative self-adjoint objects. In particular, if now $\mathbf{B} :=\mathbf{P}  \mathrm{Diag}(\mathbf{b})  \mathbf{P}^{\mathsf{T}}$ and $\mu_A$ and $\mu_B$ have zero mean, we have  $\tau(\mathbf{A} \mathbf{B} \mathbf{A} \mathbf{B}) = \tau(\mathbf{A}^2) \tau(\mathbf{B}^2)$ and more  generally crossing and non-crossing partitions contribute equally in this setting.

{\bf The high-temperature convolution -}
 A natural problem is to give a meaning for “intermediate cases”, that is to describe the spectrum of the sum of two large self-adjoint objects for cases where those objects are not commutative nor maximally non-commutative. Recently \cite{mergny_cconv}, promising progress has been made in this direction with the introduction of the so-called \emph{high-temperature convolution} (or “$c$-convolution”), which we briefly describe in this paragraph, see also \cite{benaych2021matrix}. The high-temperature convolution has been constructed by looking at spherical integrals \cite{harish1956invariant,itzykson1980planar,Marinari1994,Guionnet05,mergny2020asymptotic,benaych2011rectangular}  in a double scaling regime where $N \to \infty$ but now the usual inverse temperature parameter $\beta$ of RMT scales as $\beta_N = 2c/N$, hence the name for the convolution.
 
 The precise description of this convolution will be given later on, but for now one can think of it as  an operation taking a  parameter $c \in (0,\infty)$ and two distributions $\mu_A$ and $\mu_B$ as inputs and giving a distribution denoted by $\mu_A \oplus_c \mu_B$ as output. This high temperature convolution admits the usual and free convolution as limiting cases, since $\mu_A \oplus_{c \to 0} \mu_B \equiv \mu_A \ast \mu_B$ and $\mu_A \oplus_{c \to \infty} \mu_B \equiv \mu_A \boxplus \mu_B$. As a consequence, it forms a continuous family of convolutions interpolating between the two aforementioned ones as one varies the parameter $c$. 
 
 Let us mention that this convolution is done directly at the level of the limiting distributions and finding the corresponding 'linear algebra operation' - such that the eigenvalues of the sum of two symmetric matrices $\mathbf{A}$ and $\mathbf{B}$ with proper conditions between the two -  is an open problem.  Nevertheless, one can still develop a combinatorial formula where now the weight of a crossing partition depend on the parameter $c$, see \cite{benaych2021matrix}. If one thinks of two abstract self-adjoint objects $\mathbf{A}, \mathbf{B}$ which are '$c$-free', by which we mean that we think of $\mu_A \oplus_c \mu_B$ as the limiting spectral distribution of their sum, we have in particular (again for $\mu_A$ and $\mu_B$ with  zero mean):
\begin{equation}
\label{eq:commutativity_c}
\tau(\mathbf{A} \mathbf{B} \mathbf{A} \mathbf{B}) = \frac{\tau(\mathbf{A}^2 \mathbf{B}^2)}{c+1}    = \frac{m_2[\mu_A] m_2[\mu_B]}{c+1}\, .
\end{equation}
Thus, the high-temperature convolution corresponds in spirit to the spectrum of the sum of two large self-adjoint objects with degree of non-commutativity indexed by the parameter $c$. 

This high-temperature convolution  admits an intriguing duality with the so-called finite free convolution \cite{marcus2021polynomial,marcus2022finite}. The latter can be understood \cite{gorin2020crystallization} as the spectrum of the sum of two $\beta$-ensembles in the low-temperature limit $\beta \to \infty$, where the number of particles $N$ is fixed. This duality translates into a correspondence $ c \leftrightarrow N$ between the respective parameter of the two convolutions and can be seen as an extension for the sum of the high-low temperature duality developed in Refs.\ \cite{Duy15,Trinh19,trinh2021beta, forrester2022high}. 

In practice, the combinatorial formula developed in Ref.\ \cite{benaych2021matrix}  is too cumbersome  to compute the distribution $\mu_A \oplus_c \mu_B$, and a simpler way is to follow the road developed in Ref.\ \cite{mergny_cconv} which rely on the Markov-Krein relation \cite{kerov1998interlacing,faraut2016markov}: 
\begin{equation}
\label{eq:MK_relation}
\int_{\mathrm{Supp} \, \mu} \frac{\mathrm{d}\mu(x)}{(z-x)^c} = \exp{ \left[ - c  \int_{\mathrm{Supp} \, \nu} \mathrm{d}\nu(y) \log \left( z - y \right) \right]  } \, ,
\end{equation}
valid for any $z$ in the complex plane outside the supports of the two distributions. The distribution $\nu$ is known as the Markov-Krein transform with index $c$ (MKT$_c$)  of the distribution $\mu$ and conversely $\mu$ is the inverse Markov-Krein transform (IMKT) of $\nu$. The main reason to introduce the Markov-Krein relation is that the high-temperature convolution $\mu_A \oplus_c \mu_B$ corresponds to the classical convolution of the MKT$_c$ of $\mu_A$ and $\mu_B$. After a few simplifications developed in Ref.\ \cite{mergny_cconv}, this means that one can decompose the computation  of the high-temperature convolutions into the following steps:
\begin{enumerate}
    \item Compute the \emph{moment generating functions} (MGF) $ M_{A,B}(s) := \mathbb{E}_{ Y \sim \nu_A , \nu_B} \left[ \mathrm{e}^{s Y} \right]$ of the MKT$_c$ the two distributions $\mu_A$ and $\mu_B$. The MKT$_c$ $\nu_A$ and $\nu_B$ can be computed thanks to sophisticated integral representation developed in Ref.\ \cite{mergny_cconv}.    
    \item Compute the function
    \begin{equation}
    \label{eq:def_Uc}
        U^{(c)}(z) := \frac{1}{\Gamma(c)}  \int_{0}^{\infty} \mathrm{d}s \,  \mathrm{e}^{-zs} s^{c-1} M_{A}(s) M_B(s)  \, ,
    \end{equation}
     for $z$ high enough, that is higher than the $K = \mathrm{max}(\mathrm{Supp} \, \mu_A) +  \mathrm{max}(\mathrm{Supp} \, \mu_B)$ and then extend  analytically this function to all $z \in \mathbb{C} \setminus ( - \infty,K)$. 
     \item  Compute the \emph{Stieltjes transform} $G^{(c)}(z) := \int \mathrm{d}(\mu_A \oplus_c \mu_B)(x) \,  (z-x)^{-1}$, thanks to the formula:
     \begin{equation}
     \label{eq:Gc_Uc}
         G^{(c)}(z) := - \frac{1}{c} \frac{\mathrm{d}}{\mathrm{d}z} \log U^{(c)}(z) =  - \frac{1}{c} \frac{ (U^{(c)})'(z)}{ U^{(c)}(z)}   \, .
     \end{equation}
     \item Compute the distribution $\mu_A \oplus_c \mu_B$ thanks to the Sokochi-Plemelj formula: 
    \begin{equation}
    \label{eq:Plemelj}
        (\mu_A \oplus_c \mu_B)(x) = \frac{1}{\pi} \mathfrak{Im} \, G^{(c)}(x - \mathrm{i}0^+) \, .
    \end{equation}
\end{enumerate}
Each step can be easily approximated numerically such that one can really think of the entire process as an algorithm for computing the high temperature convolution of two distributions. 

 However, for a given choice of $\mu_A$ and $\mu_B$ and the parameter $c$, finding an \emph{explicit expression} for the density of their high-temperature convolution is  a daunting task. In fact, the only known cases where one has an explicit expression for the high temperature convolution correspond to trivial fixed points (or infinitely divisible  distributions) which, up to rescaling, are left unchanged by the high-temperature convolution. Note that  even for the free convolution, one has an  analytical expression for the density only for specific choices of the distribution $\mu_A$ and $\mu_B$ such that one should not expect to have a simple expression for the high-temperature convolution. 
 
 \section{Main result}
 The present note aims to answer this issue by providing a complete description of  $\mu_A \oplus_c \mu_B$ for a specific choice of $\mu_A$ and $\mu_B$ and \emph{any value of the parameter $c$}.  We consider the case where $\mu_A = \mu_B = \mu$, with,
 \begin{equation}
 \label{eq:symBer}
        \mu := \frac{1}{2} \delta_{-1/2} + \frac{1}{2} \delta_{1/2} \, ,
\end{equation}
since this is a famous case where the density of its free convolution with itself is known analytically and given by the (shifted and re-scaled) arscine law: for $x  \in [-1,1]$, $(\mu \boxplus \mu)(x)  = \frac{1}{\pi \sqrt{1-x^2}}$ . For the classical convolution, we have $\mu \ast \mu = \frac{1}{4} \delta_{-1} + \frac{1}{2} \delta_{0} + \frac{1}{4} \delta_{1}$, which is the re-centered binomial distribution with number of trials $n=2$ and probability of success $p=1/2$. 

{\bf Expression for the  Stieltjes transform -} Our first result writes as follows: for any $c >0$, the Stieltjes transform $ G^{(c)}$ of $\mu \oplus_c \mu$ is given by: 
\begin{equation}
\label{eq:Gc}
     G^{(c)}(z)  
     = \frac{1}{z} \frac{ {}_2F_1\left( \frac{c}{2}, 1+ \frac{c}{2}; c ; \frac{1}{z^2} \right) }{{}_2F_1 \left(\frac{c}{2} , \frac{c}{2} ; c ; \frac{1}{z^2} \right) } \, ,
\end{equation}
where ${}_2F_1(a,b;c;u) := \sum_{k=0}^{\infty} \frac{(a)_k (b)_k}{(c)_k \, k! }u^k$ is the Gauss hypergeometric function and $(a)_k := \Gamma(a+k)/\Gamma(a)$. The derivation of Eq.\ \eqref{eq:Gc} is postponed to the next section, where we also discuss how one can recover the Stietljes transform of the classical convolution and the free convolution, corresponding respectively to $c \to 0^+$ and $c \to \infty$. Using the power series of the hypergeometric functions, one gets the large $z$ behavior of the Stieltjes transform:
\begin{equation}
\label{eq:Gc_largez}
      G^{(c)}(z)  
     =  \frac{1}{z} + \frac{1}{2 z^3} + \frac{4 +3c}{8(c+1) z^5} + \frac{8 +5c}{16(c+1) z^7} +o(z^{-8})  \, ,
\end{equation}
from which we deduce that the first even moments of the symmetric distribution $\mu \oplus_c \mu$ are given by $m_2 = 1/2$, $m_4 = (4 +3c)/(8c+8)$ and $m_6 = (8 +5c)/(16c+16)$, in accordance with the first moments one can obtain with the combinatorial formula developed in Ref.\ \cite{benaych2021matrix}. 

If one introduces a new distribution  $\tilde{\rho}(x) := (\mu \oplus_c  \mu)(\sqrt{x}) \,  x^{-1/2}$ defined for any $x \in [0,1]$, then its Stieltjes transform $ \tilde{G}^{(c)}(z) := \int_0^1 \mathrm{d}\tilde{\rho}(x) \, (z-x)^{-1}$ is related to the one of $\mu \oplus_c  \mu$ by $G^{(c)}(z) =  z \tilde{G}^{(c)}(z^2)$, that is:
\begin{equation}
\label{eq:Gtilde}
      \tilde{G}^{(c)}(z)  
     =  \frac{1}{z} \frac{ {}_2F_1\left( \frac{c}{2}, 1+ \frac{c}{2}; c ; \frac{1}{z} \right) }{{}_2F_1 \left(\frac{c}{2} , \frac{c}{2} ; c ; \frac{1}{z} \right) }   \, . 
\end{equation}
Note that the change of variable  going from $\tilde{\rho}$ to  $\mu \oplus_c \mu$ admits a natural interpretation in RMT: if one think of  $\tilde{\rho}$ as the limiting spectrum of some matrix $\mathbf{M}\mathbf{M}^{\mathsf{T}}$ then $\mu \oplus_c \mu$  is the symmetrized singular value distribution of the square matrix $\mathbf{M}$, see Ref.\ \cite{BenaychGeorges2008}. Eq.\ \eqref{eq:Gtilde} expressed $\tilde{G}^{(c)}(z)$ as a product of the inverse function by the  ratio of two different hypergeometric functions, both evaluated at $1/z$. Such general form has already  appeared before in RMT in the study of the \emph{high-temperature Jacobi ensemble}, see Ref.\ \cite{trinh2021beta}. Yet, the parameters of the hypergeometric functions here are different such that - to the best knowledge of the author - the family of distributions $\mu \oplus_c \mu$ (and $\tilde{\rho}$) is a  new one in RMT. 

\begin{figure}
\includegraphics[width=8.6cm]{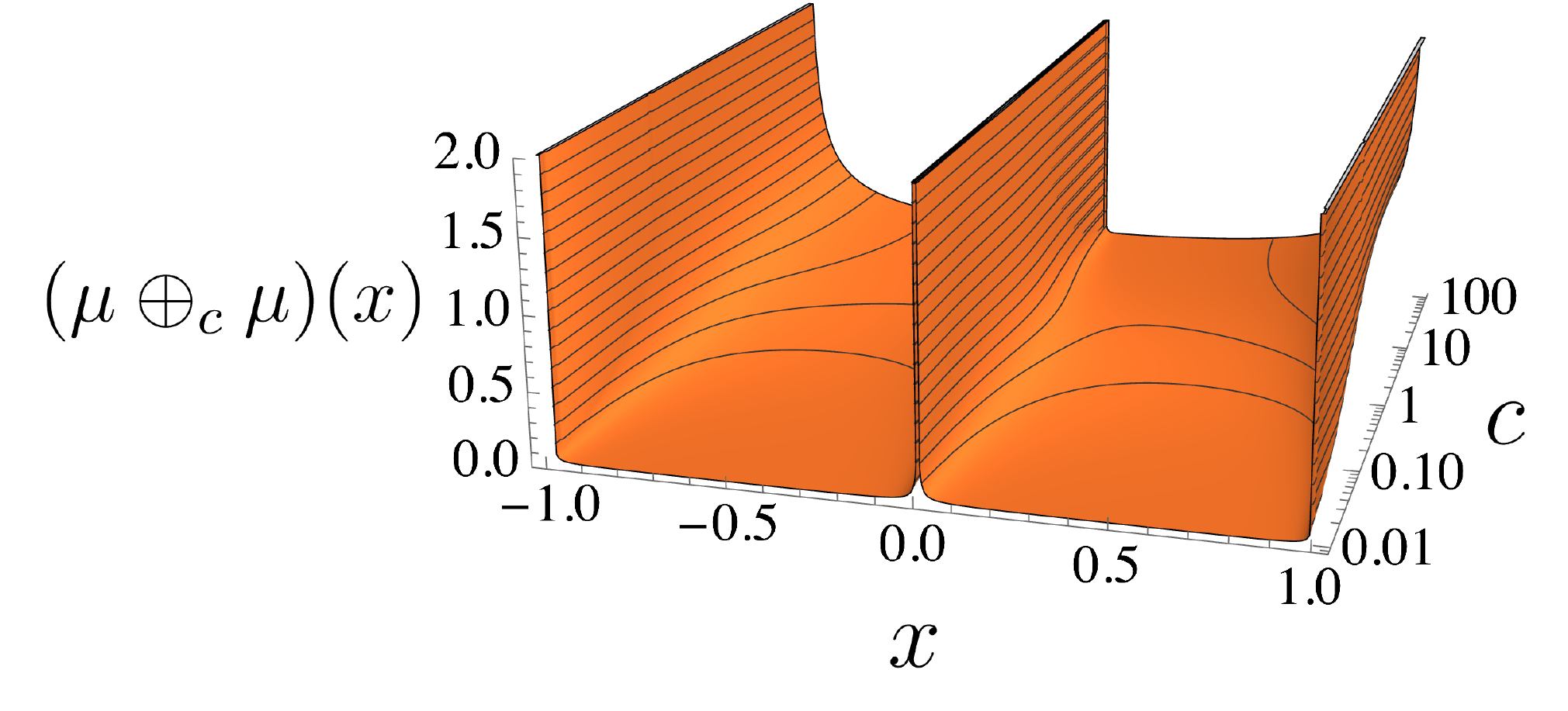}
\caption{Plot of the density $\mu \oplus_c \mu$ for $x \in [-1,1]$ and different values of the parameter $c$, in logarithmic scale. }
\label{fig:density}
\end{figure}

{\bf Expression for the density -} Our second result is written as follows:  if we define  $V_1(x) :=  {}_2F_1 \left( 1 - \frac{c}{2}, \frac{c}{2}, 1 ;x \right) $ and  $V_2(x) :=  {}_2F_1 \left( 2 - \frac{c}{2}, 1+ \frac{c}{2}, 2 ;x \right) $, then for any $c$  such that $2c \notin \mathbb{N}$, the density  $\mu \oplus_c \mu$ is given for any $x \in [-1,1]  \setminus \{-1,0,1 \}$ by: 
\begin{equation}
\begin{aligned}
\label{eq:density_allc}
   (\mu \oplus_c \mu)(x) =
   \frac{(2-c) \sin \left( c \pi/2 \right) }{2 \pi} \times \qquad \qquad \\ \frac{|x| (V_1(1{-}x^2) V_2(x^2) + V_1(x^2) V_2(1{-}x^2) )}{
   V_1(x^2)^2 + 2 \cos \left( \frac{c \pi}{2} \right)  V_1(x^2)  V_1(1{-}x^2) +  V_1(1{-}x^2)^2 } \, .  
\end{aligned}
\end{equation}
 Furthermore, one can obtain the cases where $c$ is a positive even integer by  carefully taking the limit, such that one can understand Eq.\ \eqref{eq:density_allc} as being valid for any $c>0$, after proper regularization. The distribution $ \mu \oplus_c \mu $ diverges at the points $\{ -1,0,1 \}$ and is otherwise absolutely continuous with no singular parts in $[-1,1]$. The derivation of Eq.\ \eqref{eq:density_allc} is given in the next section. A plot of the density of $\mu \oplus_c \mu$ is given in FIG. \ref{fig:density}.

 Now, for special values of the parameter $c$, this expression greatly simplifies. For example for $c{=}1$, corresponding in a sense to the mid-point between the classical and the free convolution (see Eq.\ \eqref{eq:commutativity_c}), we have:
 \begin{equation}
\label{eq:density_c1}
    (\mu\oplus_{c=1}\mu)(x)= \frac{1}{2 |x| (1{-}x^2)} \frac{1}{K(x^2)^2 + K(1{-}x^2)^2}  \, ,
 \end{equation}
 where $K(.)$ is the complete elliptic integral of the first kind, $K(u) := \int_0^{\pi/2} \mathrm{d}\theta \, (1 - u^2 \sin^2 \theta)^{-1/2}$. The expression for $c=2$ is even simpler since we have: 
 \begin{equation}
\label{eq:density_c2}
\begin{aligned}
    (\mu \oplus_{c=2} \mu)(x)= \frac{1}{|x| (1-x^2)} \frac{1}{ \left(\log \left( \frac{1 -x^2 }{x^2}  \right)  \right)^2 + \pi^2 }  \, .&&
\end{aligned}
 \end{equation} 
In practice when $c$ is an even positive integer it is easier to evaluate the Stieltjes transform thanks to Eq.\ \eqref{eq:Gc} and then use the Sokochi-Plemelj formula of Eq.\ \eqref{eq:Plemelj} rather than taking the limit in the generic expression of Eq.\ \eqref{eq:density_allc}.

 \section{Derivation of the result}
In this section we follow the steps enumerated in the previous section in order to prove Eq.\ \eqref{eq:Gc} and Eq.\ \eqref{eq:density_allc} giving the expression for the Stieltjes transform and the density, respectively. 

{\bf Computing the MGF of the MKT$_c$ -} For the symmetric Bernoulli distribution, it has been previously shown, see Refs.\ \ \cite{mergny_cconv,hjort2005exact,cifarelli1990distribution}, that the corresponding MKT$_c$ is the density of the random variable $Y' \sim \mathrm{Beta}(c/2,c/2)$.  The distribution $\mu$ is the symmetric Bernoulli distribution, shifted by $1/2$. From the Markov-Krein relation of Eq.\  \eqref{eq:MK_relation}, one sees immediately that a shift in the distribution $\mu$ introduces the same shift in the MKT$_c$. Thus, the MKT$_c$  of $\mu$ is simply the law of $Y{=} Y'{-}1/2$. From well-known properties of the Beta distribution, this means that the MGF of the MKT$_c$ of $\mu$,  $M(.) \equiv M_A(.) = M_B(.)$, is given by:
\begin{equation}
\label{eq:M.1}
    M(s) = \mathrm{e}^{s/2} \, {}_1F_1 \left( \frac{c}{2};c; s \right) \, ,
\end{equation}
where ${}_1F_1(a; b; u) := \sum_{k=0}^{\infty} \frac{(a)_k}{(b)_k} u^k$ is the \emph{confluent hypergeometric function}. Using identities \cite{DLMF} for the confluent hypergeometric, this can also be expressed in terms of  the \emph{modified Bessel function of the kind} $I_{\alpha}(.)$:
\begin{equation}
\label{eq:M.2}
    M(s) = C_1 \,  s^{(1-c)/2} I_{\frac{c-1}{2}} \left( \frac{s}{2} \right) \, ,
\end{equation}
where $ C_1 := 2 ^{c-1} \Gamma \left( \frac{c+1}{2} \right)$.

{\bf Computing the function $U^{(c)}$ -} Injecting Eq.\ \eqref{eq:M.2} into the definition of $U^{(c)}(z)$ given by Eq.\ \eqref{eq:def_Uc}, one obtains the integral representation:
\begin{equation}
    \label{eq:U_c.ex}
      U^{(c)}(z) = C_2 \,   \int_{0}^{\infty} \mathrm{d}s \,  \mathrm{e}^{-zs}  \left( I_{\frac{c-1}{2}} \left( s/2 \right) \right)^2 \, ,
\end{equation}
where  $ C_2 := C_1^2/\Gamma(c)$ is a constant that will not contribute to the expression of the Stietljes transform (and hence the density). The square of the Bessel function can be expressed as an integrated Bessel function thanks to the formula  \cite{DLMF}: 
\begin{equation}
\label{eq:square_I}
    \left( I_{\frac{c-1}{2}} \left( s/2 \right) \right)^2 = \frac{2}{\pi} \int_0^{\frac{\pi}{2}} \mathrm{d}\theta \,  I_{c-1} \left( s \cos \theta \right) \, .
\end{equation}
If we do the change of variable $s \to s  \cos \theta$ in Eq.\ \eqref{eq:U_c.ex} and then $\theta \to \mathrm{arcos} \left( \mathrm{cosh} \, \theta \right)$ we have:
\begin{equation}
\label{eq:Uc.ex.2}
    U^{(c)}(z)  = C_3  \int_0^{\infty} \mathrm{d}s \,  I_{c-1} \left( s \right) \left(\int_0^{\infty} \mathrm{d}\theta \, \mathrm{e}^{-(zs) \, \mathrm{cosh} \, \theta} \right)  \, ,
\end{equation}
with $C_3 = 2 \, C_2 / \pi$. The integral with respect to  the variable $\theta$ is the integral representation \cite{DLMF} of the \emph{Bessel function of the second kind} $K_0(.)$, such that we have:
\begin{equation}
\label{eq:Uc.ex.3}
    U^{(c)}(z) = C_3 \,   \int_0^{\infty} \mathrm{d}s \,  I_{c-1} \left( s \right) K_0( zs) \, .
\end{equation}
By identities for the integral of the product of two Bessel functions of different kinds, see  \cite{DLMF}, one can finally express $ U^{(c)}$ in terms of a hypergeometric function: 
\begin{equation}
    \label{eq:U_c.ex.4}
      U^{(c)}(z) = \Gamma(c) z^{-c} \, {}_2F_1 \left( \frac{c}{2}, \frac{c}{2};c; \frac{1}{z^2} \right)   \,  .
\end{equation}

{\bf Computing the Stieltjes transform -} In order to compute $G^{(c)}$  given by Eq.\ \eqref{eq:Gc_Uc}, we  first need to compute the derivative of $U^{(c)}(z)$. Using the differentiation formula  \cite{DLMF} for the hypergeometric function, $ (\mathrm{d}/\mathrm{d}u) \,  \,  {}_2F_1 \left( a, b ;c; u \right) = (ab/c)  \, \, {}_2F_1 \left( a+1, b+1 ;c+1; u \right)$, we get: 
\begin{equation}
\begin{aligned}
    \label{eq:DU_c}
      &(U^{(c)})'(z) 
     = \Gamma(c) (-c) z^{-c-1} \biggl[  \frac{c}{2 z^2}  \times   \\
      & \, {}_2F_1 \left(  1+ \frac{c}{2}, 1 + \frac{c}{2}; 1 + c; \frac{1}{z^2} \right) +   {}_2F_1 \left( \frac{c}{2}, \frac{c}{2};c; \frac{1}{z^2} \right)  \biggr] \, .
\end{aligned}
\end{equation}
 Next, using identities  \cite{DLMF}  between \emph{contiguous} hypergeometric functions, the sum inside the brackets simplifies such that the derivative of  $U^{(c)}(z)$ writes:
\begin{equation}
    \label{eq:DU_c.2}
      (U^{(c)})'(z)  = \Gamma(c) (-c) z^{-c} \, \frac{ {}_2F_1 \left( \frac{c}{2}, 1+\frac{c}{2};c; \frac{1}{z^2} \right)}{z}   \, .
\end{equation}
 Injecting the expression of $U^{(c)}(z)$ and its derivative, given respectively by Eq.\ \eqref{eq:U_c.ex.4} and Eq.\ \eqref{eq:DU_c.2}, in Eq.\ \eqref{eq:Gc_Uc} gives the desired expression for the Stieltjes transform, see Eq.\ \eqref{eq:Gc}. 

The limiting case $c \to 0^+$ given by the classical convolution can be recovered by using an expansion for small $c$ in the hypergeometric functions entering the expression of $G^{(c)}$. For the numerator, we get: 
\begin{equation}
    \label{eq:num_smallc}
    {}_2F_1 \left( \frac{c}{2}, 1+\frac{c}{2};c; \frac{1}{z^2} \right) =  1 + \sum_{k=1}^{\infty} ( \frac{1}{2} + o_c(1)) \frac{1}{z^{2k}}  \, ,
\end{equation}
that is:
\begin{equation}
    \label{eq:num_smallc.2}
    {}_2F_1 \left( \frac{c}{2}, 1+\frac{c}{2};c; \frac{1}{z^2} \right) =  1 + \frac{1}{(z^2-1)} +  o_c(1) \, .
\end{equation}
Similarly, the hypergeometric function in the denominator is equal to $1 +o_c(1)$. Combining these two asymptotic behaviors, we get for the Stietljes:
\begin{equation}
    \label{eq:Gc_smallc.1}
   G^{(c \to 0^{+})} = \frac{1}{z} + \frac{1}{2z(z^2 -1)} \, ,
\end{equation}
which decomposes into simple elements as: 
\begin{equation}
    \label{eq:Gc_smallc.2}
   G^{(c \to 0^{+})} = \frac{1}{4(z+1)} + \frac{1}{2z} + \frac{1}{4(z-1)}  \, ,
\end{equation}
as expected for the Stieltjes transform of the centered binomial distribution, $\mu \ast \mu = \frac{1}{4}  \delta_{-1} + \frac{1}{2} \delta_0 + \frac{1}{4} \delta_{1}$. 

 The limiting case $c \to \infty$ corresponding to the free convolution  requires more work, and we only sketch the main ingredients to recover the Stieltjes transform of the arcsine law. The idea is to use the integral representation \cite{DLMF} of the hypergeometric function:
  \begin{equation}
  \begin{aligned}
    \label{eq:int_rep_HF}
 {}_2F_1 \left( a ,b ;c; u \right)= C_3 & \int_0^1 \mathrm{d}t \, t^{b-1} \\
 & \times  (1-t)^{c-b-1} (1-tu)^{-a}    \, , 
  \end{aligned}
\end{equation}
 with $ C_4 = \frac{\Gamma(c)}{\Gamma(c-b) \Gamma(b)}$ and $c>b>0$. If we denote by $F_{\eta}(c,u) := {}_2F_1 \left( c/2 , \eta + c/2 ;c; u \right)$ with  $\eta=1$ for the hypergeometric function in the numerator of Eq.\ \eqref{eq:Gc} and $\eta=0$ for the denominator, this means that we can write  $F_{\eta}(c,u)$ as:
 \begin{equation}
    \label{eq:Fbc}
  F_{\eta}(c,u) \propto  \int_0^1 \mathrm{d}t \,  \mathrm{e}^{\frac{c}{2} g(t,u)} h(t)   \, ,
\end{equation}
with $g(t,z):= \log(t(1-t)) - \log(1-tu)$ and $h(t):=t^{\eta-1}(1-t)^{1-\eta}$. As $c\to \infty$, Eq.\ \eqref{eq:Fbc} can be approximated by Laplace method and the results writes: 
 \begin{equation}
\label{eq:Fbc.2}
  F_{\eta}(c,u) \underset{c \to \infty}{\sim} K_c(u) (1-u)^{-\eta/2}   \, ,
\end{equation}
where $ K_c(u)$ is a function independent of the parameter ${\eta}$. Thus, if we inject this asymptotic behavior in Eq.\ \eqref{eq:Gc} we get: 
\begin{equation}
    \label{eq:Gc_largec.1}
   G^{(c \to \infty)} =  \frac{1}{z} \frac{1}{\sqrt{1 - \frac{1}{z^2}}}  \, ,
\end{equation}
which is indeed the Stietljes transform of $\mu \boxplus \mu$, see \cite{potters-bouchaud}.

{\bf Computing the density -} The explicit expression for the density is obtained thanks to the Sokochi-Plemelj formula of Eq.\ \eqref{eq:Plemelj} and the expression of Eq.\ \eqref{eq:Gc} for the Stietljes transform, by looking carefully at the behavior of the hypergeometric functions near their branch cuts. As the situation is very similar for both the numerator and denominator, we only detail the complete computation for the latter case. The idea is to use both the integral representation of Eq.\ \eqref{eq:int_rep_HF} for $a=c/2$ and $b = \eta +c/2$ and the behavior of the power function near its branch cut. For $u>0$, we have $( u +\mathrm{i}0^+)^{\alpha} = u^{\alpha}$ but otherwise:
\begin{equation}
\label{eq:powerfunction_bc}
    ( -u +\mathrm{i}0^+)^{\alpha} = \cos( \pi \alpha)  |u|^{\alpha} + \mathrm{i} \sin( \pi \alpha)   |u|^{\alpha} \, . 
\end{equation}
As $z \to x - \mathrm{i} 0^+$ with $x \in [-1,1]$, we have
\begin{equation}
    (1{-}t/z^2)^{-c/2} {=} (1{-}t/x^2 {+} \mathrm{i} \,  \mathrm{sign}(x) \mathrm{0}^+)^{-c/2} \, ,
\end{equation}
 such that we need to differentiate the cases $t<x^2$  and $t>x^2$ in Eq.\ \eqref{eq:int_rep_HF}. Since  the distribution $\mu \oplus_c \mu$ is symmetric, we also fix $x>0$. Thus, if we introduce the two functions $J_{1,2}(x)$ corresponding respectively to the split of the integral of Eq.\ \eqref{eq:int_rep_HF} for $a=c/2$ and $b=c/2$, into the segment $[0,x^2]$ and $[x^2,1]$:
 \begin{equation}
    \label{eq:def_J1}
      J_1(x) :=  C_4 \int_{0}^{x^2} \mathrm{d}t \,  (t(1-t))^{-c/2} \left(1- \frac{t}{x^2}\right)^{-c/2} \, ,
\end{equation}
and 
\begin{equation}
    \label{eq:def_J2}
J_2(x) :=  C_4 \int_{x^2}^{1} \mathrm{d}t \,  (t(1-t))^{-c/2} \left(\frac{t}{x^2} -1 \right)^{-c/2} \, , 
\end{equation}
then  the real and imaginary parts of the hypergeometric function in the denominator of Eq.\ \eqref{eq:Gc} are given by:
\begin{equation}
    \label{eq:Re_2F1}
      \mathfrak{Re}  \,  {}_2F_1 \left( \frac{c}{2}, \frac{c}{2};c; \frac{1}{(x{-} \mathrm{i}0^+)^2} \right) = J_1(x) + \cos \left( \frac{\pi c}{2} \right) J_2(x) \, ,
\end{equation}
and 
\begin{equation}
    \label{eq:Im_2F1}
         \mathfrak{Im}  \,  {}_2F_1 \left( \frac{c}{2}, \frac{c}{2};c; \frac{1}{(x{-} \mathrm{i}0^+)^2} \right) = - \sin \left( \frac{\pi c}{2} \right) J_2(x) \, .
\end{equation}
If we now perform the change of variable $t \to x^2 t$ in Eq.\ \eqref{eq:def_J1}, we can rewrite $J_1(x)$ as: 
 \begin{equation}
    \label{eq:J1.1}
      J_1(x) =  C_4 \,  x^{c} \int_{0}^{1} \mathrm{d}t \, t^{c/2-1} \left(1- t\right)^{-c/2} \left(1- x^2 t\right)^{-c/2 -1}   \,  .
\end{equation}
By Eq.\ \eqref{eq:int_rep_HF} and up to a multiplicative constant we recognize the integral in Eq.\ \eqref{eq:J1.1} as the hypergeometric function $V_1(x) :=  {}_2F_1 \left( 1 - \frac{c}{2}, \frac{c}{2}, 1 ;x \right)$. The multiplicative constant can be simplified thanks to the complement formula of the Gamma function $\Gamma(1-z)\Gamma(z) = \pi/\sin(\pi z)$ for $z \in \mathbb{C} \setminus \mathbb{Z}$, and we finally obtain: 
 \begin{equation}
    \label{eq:J1.2}
      J_1(x) =  \frac{\pi \Gamma(c)}{\Gamma \left( \frac{c}{2}\right)^2 \sin \left( \frac{c \pi}{2} \right)}  x^{c} \,  V_1(x^2)   \, .
\end{equation}
Note that the integral representation of Eq.\ \eqref{eq:J1.1} is actually only valid for $c \in (0,2)$ but by  analytic continuation of the hypergeometric function, the result holds for any $c>0$ such that $2c \notin \mathbb{N}$, due to the presence of the inverse of the sinus function in Eq.\ \eqref{eq:J1.2}. 

Similarly, by  the change of variable $t \to x^2 +(1-x^2) t^2$ in Eq.\ \eqref{eq:def_J2}, $J_2(x)$ can be expressed as: 
 \begin{equation}
    \label{eq:J1.3}
      J_2(x) =  \frac{\pi \Gamma(c)}{\Gamma \left( \frac{c}{2}\right)^2 \sin \left( \frac{c \pi}{2} \right)}  x^{c} \,  V_1(1 -x^2)   \, .
\end{equation}
Thanks to Eq.\ \eqref{eq:Re_2F1} and Eq.\ \eqref{eq:Im_2F1}, one has the complete behavior near the branch cut for the  denominator of Eq.\ \eqref{eq:Gc}.

We then sketch the remaining steps to get the analytical expression for density: one can then repeat the exact same previous computation for the numerator of Eq.\ \eqref{eq:Gc}, where instead of the function $ V_1(.)$, the function $V_2(.)$ will appear (with a different multiplicative constant) when splitting the integral into the segments $[0,x^2]$ and $[x^2,1]$. All in all, one gets the density by taking the imaginary part of the entire expression, divided by $\pi$. After simplification with the trigonometric identity $\cos(c \pi/2)^2 {+} \sin(c \pi/2)^2 {=}1$, appearing when one multiplies the denominator of Eq.\ \eqref{eq:Gc} by its conjugate, one gets the desired expression of Eq.\ \eqref{eq:density_allc} for the density.

\section{Conclusion}

In this note, we studied the high-temperature convolution introduced in Ref \cite{mergny_cconv}, between two symmetric Bernoulli distributions. Our result provides a new family of distribution, indexed by the parameter $c$ of the high-temperature convolution, interpolating between the (shifted) binomial distribution with parameter $(2,1/2)$ and the (shifted and re-scaled) arcsine law. This family of distribution constitutes the first non-trivial case for the analytical expression of the high-temperature,  and we believe that the ideas developed in this note can be used to obtain the density of the high-temperature convolution in other cases. The obtained distribution $\mu \oplus_c \mu$ is absolutely continuous between each singular points (here being given by $\{-1,0,1\}$) of the classical convolution  $\mu \ast \mu$, and we conjectured this phenomenon to be a specific feature of the high-temperature convolution. Our result paves the way for a better understanding of this new convolution and can serve as a benchmark for future construction of the underlying linear algebra operation. 

\section{Acknowledgments}

The author would like to warmly thank M. Potters and J-P. Bouchaud for fruitful discussions concerning the high-temperature convolution and precious comments regarding this note. 

\bibliographystyle{apsrev4-1}
\bibliography{biblio}

\end{document}